\documentstyle[amssymb,12pt]{article}

\relax
\def\frac#1#2{{#1\over#2}}
\def\<{\langle}\def\>{\rangle}
\newcount\parenthesis \parenthesis=0 \newcount\n
\def\({\global\advance\parenthesis by1\left(}
\def\){\global\advance\parenthesis by-1\right)}
\def\[{\relax} 
\def\]{\relax} 
\def\Loop#1\Repeat{\global\n=0\global\let\body=#1\iterate}
\def\iterate{\body\let\next=\iterate\else\let\next=\relax\fi\next}
\def\ldd{\ifnum\n<\parenthesis\global\advance\n by1
\left.\nulldelimiterspace=0pt\mathsurround=0pt}
\def\rdd{\ifnum\n<\parenthesis\global\advance\n by1
\right.\nulldelimiterspace=0pt\mathsurround=0pt}
\def\nl{\Loop\rdd\Repeat\hfill\cr\qdd\Loop\ldd\Repeat{}}
\def\off#1{\dimen0=#1sp\divide\dimen0 by 3\hskip\dimen0\relax}
\def\Nl{\hfill\cr}
\def\qdd{\quad\quad}

\newcommand{\ov}[2]%
    {%
     \begin{array}[b]{@{}c@{}} {\scriptscriptstyle #2} \\ #1 \end{array}}

\newcommand{\om}[1]{\ov{\Omega}{(#1)}}
\newcommand{\TT}[1]{\ov{T}{(#1)}}

\newcommand{\ghp}{{\cal P}}
\newcommand{\gh}{\mbox{gh}}
\newcommand{\antigh}{\mbox{antigh}}

\newcommand{\Halmos}{\hfill\rule{.6 em}{.6 em}}
\newtheorem{theorem}{Theorem}

\newenvironment{proof}{\noindent {\bf Proof } }{\Halmos}

\title{BRST Structure of Polynomial Poisson Algebras}
\author{Alain Dresse \and Marc Henneaux\thanks%
{Also at Centro de Estudios Cient\'{\i}ficos de Santiago, Casilla 16443,
Santiago 9, Chile} \\%
Facult\'e des Sciences, Universit\'e Libre de Bruxelles, \\
Campus Plaine C.P. 231, B-1050 Bruxelles (Belgium)}

\begin{document}

\maketitle

\abstract{%
The BRST structure of polynomial Poisson algebras is investigated. It
is shown that Poisson algebras provide non trivial models where the
full BRST recursive procedure is needed. Quadratic Poisson algebras
may already be of arbitrarily high rank. Explicit examples are
provided, for which the first terms of the BRST generator are
given. The calculations are cumbersome but purely algorithmic, and
have been treated by means of the computer algebra system REDUCE. Our
analysis is classical ($=$ non quantum) throughout.}

\section{Introduction}
Polynomial algebras with a Lie bracket fulfilling the derivation
property
\begin{equation}
[f g, h] = f [g,h] + [f,h]g
\end{equation}
are called polynomial Poisson algebras and play an increasingly
important role in various areas of theoretical physics \cite{Nak:,%
Pri:,Skl:,Zam:,FatZam:,Oh:,TarTakFad:,BakMat:,BhaRam:,GraZhe:}.  In
terms of a set of independent generators $G_a$, $a = 1, \ldots, n$,
the brackets are given by
\begin{equation}\label{basic_bracket}
[G_a, G_b] = C_{ab}(G)
\end{equation}
where $C_{ab} = - C_{ba}$ are polynomials in the $G$'s%
\footnote{We shall restrict here the analysis to ordinary polynomial
algebras with commuting generators, but one can easily extend the study
to the graded case with both commuting and anticommuting
generators.}.  If the polynomials $C_{ab}(G)$ vanish when the $G$'s
are set equal to zero, i.e. if they have no constant part, the
polynomial algebra is said to be first class, in analogy with the
terminology for constrained Hamiltonian systems (see
e.g. \cite{HenTei:QuaGauSys}).  An important class of first class
Poisson algebras are symmetric algebras over a finite dimensional Lie
algebra. In that case, the bracket (\ref{basic_bracket}) belongs to
the linear span of the $G_a$'s, i.e. the $C_{ab}(G)$ are homogeneous
of degree one in the $G$'s, $[G_a, G_b] = C_{ab}{}^c G_c$. We shall
call this situation the ``Lie algebra case'', and refer to the non Lie
algebra case as the ``open algebra case'' using again terminology from
the theory of first class constrained systems \cite{HenTei:QuaGauSys}%
\footnote{
It should be stressed that the polynomial algebra generated by the
$G$'s, equipped with the bracket (\ref{basic_bracket}) is {\em always} an
infinite-dimensional Lie algebra, even in the ``open algebra''
case.}.

The purpose of this paper is to investigate the BRST structure of
first class Poisson algebras. The BRST formalism has turned out
recently to be the arena of a fruitful interplay between physics and
mathematics (see e.g. \cite{HenTei:QuaGauSys} and references
therein).  A crucial ingredient of BRST theory is the recursive
pattern of homological perturbation theory \cite{Sta:} which allows
one to construct the BRST generator step by step. In most
applications, however, this recursive construction collapses almost
immediately, and, to our knowledge, no example has been given so far
for which the full BRST machinery is required (apart from the
field-theoretical membrane models \cite{Hen:PhyLet,FujKub:}). We show
in this paper that Poisson algebras---actually, already quadratic
Poisson algebras---offer splendid examples illustrating the complexity
of the BRST construction. While Lie algebras yield a BRST generator of
rank 1 (see e.g. \cite{HenTei:QuaGauSys}), the BRST charge for
quadratic Poisson algebras can be of arbitrarily high rank. We also
point out that BRST concepts provide intrinsic characterizations of
Poisson algebras.

In the next section, we briefly review the BRST construction. We then
discuss how it applies to Poisson algebras, even when the generators
$G_a$ are not realized as phase space functions of some dynamical
system. We analyze the BRST cohomology and introduce the concepts of
covariant and minimal ranks, for which an elementary theorem is
proven.  Quadratic algebras are then shown to provide models with
arbitrarily high rank. These contain ``self-reproducing'' algebras for
which the bracket of $G_a$ with $G_b$ is proportional to the product
$G_a G_b$. The first few terms in the BRST generator are also computed
for more general algebras by means of a program written in REDUCE. The
paper ends with some concluding remarks on the quantum case.

\section{A Brief Survey of the BRST Formalism}
We follow the presentation of \cite{HenTei:QuaGauSys}, to which we
refer for details and proofs.
Given a set of independent functions $G_a(q,p)$ defined in some phase
space $P$ with local coordinates $(q^i, p_i)$ and fulfilling the first
class property $[G_a, G_b] \approx 0$, where $\approx$ denotes equality
on the surface $G_a(q,p) = 0$, one can introduce an odd generator
$\Omega$ (``the BRST generator'') in an extended phase space
containing further fermionic conjugate pairs $(\eta^a, \ghp_a)$ (the
``ghost pairs'') which has the following properties :
\begin{eqnarray}
[\Omega, \Omega] &=& 0 \label{nilpotency} \\
\Omega &=& G_a \eta^a + \mbox{``more''}.
\end{eqnarray}
Here, ``more'' stands for terms containing at least one ghost momentum
$\ghp_a$. We take the ghosts $\eta^a$ to be real and their momenta
imaginary, with graded Poisson bracket
\begin{equation}
[\ghp_a, \eta^b] = - \delta_a{}^b
\end{equation}

The BRST derivation $s$ in the extended phase space is generated by
$\Omega$,
\begin{equation}
s \bullet = [ \bullet, \Omega]
\end{equation}
and is a differential ($s^2 = 0$) because of (\ref{nilpotency}). One also
introduces a grading, the ``ghost number'' by setting
\begin{equation}
\gh \eta^a = - \gh \ghp_a = 1, \quad \gh q^i = \gh p_i = 0.
\end{equation}
The ghost number of the BRST generator is equal to 1.

The BRST generator $\Omega$ is constructed recursively as follows. One
sets
\begin{equation}
\Omega = \om{0} + \om{1} + \cdots
\end{equation}
where $\om{k}$ contains $k$ ghost momenta. One has
$\om{0} = G_a \eta^a$. The nilpotency condition becomes,
in terms of $\om{k}$,
\begin{equation}\label{delta-om=d}
\delta \om{p+1} + \ov{D}{(p)} = 0
\end{equation}
where $\ov{D}{(p)}$ involves only the lower order
$\om{s}$ with $s \leq p$ and is defined by
\begin{equation}\label{d-p}
\ov{D}{(p)} = 1 / 2 \left[
  \sum^p_{k=0} [\om{k} , \om{p-k}]_{\mbox{orig}} +
  \sum^{p-1}_{k=0} [ \om{k+1} ,
  \om{p-k}]_{\ghp, \eta}
\right].
\end{equation}

Here, the bracket $[\; , \;]_{\mbox{orig}}$ refers to the Poisson
bracket in the original phase space, which only acts on the $q^i$ and
$p_i$, and not on the ghosts, whereas $[\; , \;]_{\ghp, \eta}$ refers
to the Poisson bracket acting only on the ghost and ghost momenta
arguments and not on the original phase space variables. The
``Koszul'' differential $\delta$ in (\ref{delta-om=d}) is defined by
\begin{equation}\label{koszul}
\delta q^i = \delta p_i = 0, \quad \delta \eta^a = 0, \quad \delta
\ghp_a = - G_a
\end{equation}
and is extended to arbitrary functions on the extended phase space as
a derivation. One easily verifies that $\delta^2 = 0$.

Given $\om{s}$ with $s \leq p$, one solves (\ref{delta-om=d}) for
$\om{p+1}$. This can always be done because $\delta
\ov{D}{(p)} = 0$, and because $\delta$ is acyclic in positive degree.
One then goes on to $\om{p+2}$ etc... until one reaches the complete
expression for $\Omega$. The last function $\om{k}$ that can be non
zero is $\om{n-1}$ where $n$ is the number of constraints. Indeed, the
product $\eta^{a_1} \cdots \eta^{a_n} \eta^{a_{n+1}}$ of $n+1$
anticommuting ghost variables in $\ov{\Omega}{(n)}$ is zero.  The
function $\ov{\Omega}{(p+1)}$ is determined by (\ref{delta-om=d}) up
to a $\delta$-exact term. This amounts to making a canonical
transformation in the extended phase space.

\section{First Class Polynomial Poisson algebras}

The standard BRST construction recalled in the previous section
assumes that the $G_a$'s are realized as functions on some phase
space, and allows the $C^c{}_{ab}$ in
\begin{equation}
[G_a, G_b] = C^c{}_{ab} G_c
\end{equation}
to be functions of $q^i$ and $p_i$. However, when the $C^c{}_{ab}$'s
depend on the $q$'s and $p$'s only through the $G_a$'s themselves, as
is the case when the $G_a$'s form a first class polynomial Poisson
algebra, one can define the BRST generator directly in the algebra
$\Bbb{C}\,(\ghp_a) \otimes \Bbb{C}\,(G_a) \otimes \Bbb{C}\,(\eta^a)$
of polynomials in the $G$'s, the $\eta$'s and the $\ghp$'s without any
reference to the explicit realization of the $G$'s as phase space
functions%
\footnote{
In agreement with the notations of \cite{HenTei:QuaGauSys}, we denote
the algebra of polynomials in the anticommuting variables $\ghp_a$
with complex coefficients by $\Bbb{C}\,(\ghp_a)$, and not by the more
familiar notation $\Lambda(\ghp_a)$. A typical element of
$\Bbb{C}\,(\ghp_1)$ is $a + b \ghp_1$ with $a, b \in \Bbb{C}\,$ since
$(\ghp_1)^2 = 0$.  }.  That is, the BRST generator can be associated
with the Poisson algebra itself.

The reason for which this can be done is that both the Koszul
differential $\delta$ defined by (\ref{koszul}) {\em and}
the $\ov{D}{(p)}$
in (\ref{d-p}) involve only $G_a$ and not $q^i$ or $p_i$ individually.
Thus, $\ov{\Omega}{(p+1)}$ can be taken to depend only on $G_a$. The
BRST generator is defined accordingly in the algebra $\Bbb{C}\,(\ghp_a)
\otimes \Bbb{C}\,(G_a) \otimes \Bbb{C}\,(\eta_a)$.

One can give an explicit solution of (\ref{delta-om=d}) in terms of the
homotopy $\sigma$ defined on the generators by
\begin{equation}
\sigma G_a = - \ghp_a, \quad \sigma \ghp_a = \sigma G_a = 0
\end{equation}
and extended to the algebra $\Bbb{C}\,(\ghp_a) \otimes \Bbb{C}\,(G_a) \otimes
\Bbb{C}\,(\eta_a)$ as a derivation,
\begin{equation}\label{sigma}
\sigma = - \ghp_a \frac{\partial}{\partial G_a}.
\end{equation}
One has
\begin{equation}
\sigma \delta + \delta \sigma = N
\end{equation}
where $N$ counts the degree in the $G$'s and the $\ghp$'s. Hence, if
$\ov{D_m}{(p)}$ is the term of degree $m$ in $(G, \ghp)$ of
$\ov{D}{(p)}$, a solution of (\ref{delta-om=d}) is given by
\begin{equation}\label{om-p}
\om{p+1} = - \sum_m 1/m \left( \sigma \ov{D}{(p)}_m
\right)
\end{equation}
since $\delta \ov{D}{(p)}=0$ \cite{HenTei:QuaGauSys} and $m >
0$ (one has $m \geq p$ and for $p=0$, $m \geq 1$ because $[\om{0},
\om{0}]$ contains $G_a$ by the first class property).

It should be stressed that the partial derivations $\partial/\partial
G_a$ in(\ref{sigma}) are well defined because the functions on which
they act depend only on on $G_a$. For an arbitrary function of $q^i,
p_i$, $\partial F / \partial G_a$ would not be well defined even if
the constraints $G_a$ are independent (i.e. irreducible) as here. One
must specify what is kept fixed. For example, if there is one
constraint $p_1 = 0$ on the four-dimensional phase space $(q^1, p_1),
(q^2, p_2)$, then $\partial p_2 / \partial p_1 = 0$ if one keeps $q^1,
q^2$ and $p_2$ fixed, but $\partial p_2 / \partial p_1 = 1$ if one
keeps $q^1, q^2$ and $p_2 - p_1$ fixed. Note that the subsequent
developments require only that the $C^a{}_{b c}$ be functions of the
$G_a$, but not that these functions be polynomials. We consider here
the polynomial case for the sole sake of simplicity.

As mentioned earlier, the solution (\ref{om-p}) of the equation
(\ref{delta-om=d}) is not unique. We call it the ``covariant
solution'' because the homotopy $\sigma$ defined by (\ref{sigma}) is
invariant under linear redefinitions of the generators.

\noindent{\bf Example:} for a Lie algebra
\begin{equation}
[G_a, G_b] = C^c{}_{ab} G_c
\end{equation}
the covariant BRST generator is given by
\begin{equation}\label{L-A-omega}
\Omega = G_a \eta^a - 1/2 \ghp_a C^a{}_{bc} \eta^c \eta^b.
\end{equation}
Its nilpotency expresses the Jacobi identity for the structure
constants $C^a{}_{bc}$. One has $\om{p} = 0$ for $p \geq
2$.

In general, the BRST generator $\Omega$ for a generic Poisson algebra
contains higher order terms whose calculation may be quite cumbersome.
However, because the procedure is purely algorithmic, it can be
performed by means of an algebraic program like REDUCE.

The cohomology of the Poisson algebra may be defined to be the
cohomology of the BRST differential $s$ in the algebra $\Bbb{C}\,(\ghp_a)
\otimes \Bbb{C}\,(G_a) \otimes \Bbb{C}\,(\eta_a)$. Because $s$ contains
$\delta$ as its piece of lowest antighost number (with
$\antigh(\ghp_a) = 1, \antigh(\mbox{anything else}) = 0$), and because
$\delta$ provides a resolution of the zero-dimensional point $G_a =
0$, standard arguments show that the cohomology of $s$ is isomorphic
to the cohomology of the differential $s'$ in $\Bbb{C}\,(\eta^a)$,
\begin{equation}\label{eq:sPrime}
s' \eta^a = 1/2 C^a{}_{bc} \eta^b \eta^c
\end{equation}
where $C^a{}_{bc}$ is defined by
\begin{equation}
C^a{}_{bc} = \left.\frac{\partial C_{bc}}{\partial G_a}\right|_{G = 0}
\end{equation}

The $C^a{}_{bc}$ fulfill the Jacobi identity so that $s'^2 = 0$. Hence,
they are the structure constants of a Lie algebra, which is called the
Lie algebra underlying the given Poisson algebra.

Because of (\ref{eq:sPrime}), the BRST cohomology of a Poisson algebra
is isomorphic to the cohomology of the underlying Lie algebra. For a
different and more thorough treatment of Poisson cohomology, see
\cite{Hue:}.

\section{Rank}

Again in analogy with the terminology used in the theory of
constrained systems, we shall call {\em ``covariant rank''} of a first
class polynomial Poisson algebra the degree in $\ghp_a$ of the
covariant BRST generator. This concept is invariant under linear
redefinitions of the generators because the covariant BRST generator
is itself invariant if one transforms the ghosts and their momenta as
\begin{eqnarray}
G_a &\rightarrow& \bar{G}_a = A_a{}^b G_b \\
\ghp_a &\rightarrow& \bar{\ghp}_a = A_a{}^b \ghp_b \\
\eta^a &\rightarrow& \bar{\eta}^a = (A^{-1})_b{}^a \eta^b
\end{eqnarray}

We shall call {\em ``minimal rank''} the degree in $\ghp_a$ of the
solution of $[ \Omega, \Omega] = 0$ of lowest degree in $\ghp$ (i.e.,
one chooses at each stage $\om{p+1}$ in such a way that $\Omega$ has
lowest possible degree in $\ghp$). It is easy to see that for a Lie
algebra, the concepts of covariant and minimal ranks coincide. As we
shall see on an explicit example below, they do not in the general
case.

Now, for a Lie algebra, the rank is not particularly interesting in
the sense that it does not tell much about the structure of the
algebra : the rank of a Lie algebra is equal to zero if and only if
the algebra is abelian. It is equal to one otherwise. For non linear
Poisson algebras, the rank is more useful. All values of the rank
compatible with the trivial inequality
\begin{equation}
rank \leq n-1
\end{equation}
may occur. Thus, the rank of the BRST generator provides a non trivial
characterization of Poisson algebras. Conversely, non linear Poisson
algebras yield an interesting illustration of the full BRST machinery
where higher order terms besides $\om{1}$ are required in $\Omega$ to
achieve nilpotency.

\section{Upper bound on the rank}

One can understand the fact that the rank of a Lie algebra is at most
equal to one by introducing a degree in $\Bbb{C}\,(\ghp_a) \otimes
\Bbb{C}\,(G_a) \otimes \Bbb{C}\,(\eta_a)$ different from the ghost degree as
follows.

\begin{theorem}
Assume that one can assign a ``degree'' $n_a \geq 1$ to the
generators $G_a$ in such a way that the bracket decreases the degree by
at least one,
\begin{equation}\label{deg-g=n}
\deg G_a = n_a, \; \deg([G_a, G_b]) \leq n_a + n_b - 1.
\end{equation}
Then, one can bound the covariant and minimal ranks of the algebra by
$\sum_a (n_a - 1) + 1$,
\begin{equation}
r \leq \sum_a (n_a - 1) + 1
\end{equation}
\end{theorem}

In the case of a Lie algebra, one can take $n_a = 1$ for all the
generators since $\deg([G_a, G_b])$ is then equal to one and fulfills
(\ref{deg-g=n}). The theorem then states that the rank is bounded by one, in
agreement with (\ref{L-A-omega}).

\begin{proof}
Assign the following degrees to $\eta^a$ and $\ghp_a$,
\begin{equation}
\deg \eta^a = - n_a + 1, \; \deg \ghp_a = n_a - 1
\end{equation}
If $\delta A = B$ and $\deg B = b$, then $\deg A = b - 1$ since
$\delta$ increases the degree by one. Now $\om{0} = G_a
\eta^a$ is of degree one. It follows that $[\om{0},
\om{0}] = [\om{0},\om{0}]_{\mbox{orig}}$ is of degree $\leq 1$ and
hence, by (\ref{delta-om=d}) and (\ref{d-p}), $\deg \ov{\Omega}{(1)}
\leq 0$. More generally, one has $\deg \om{k} \leq -k + 1$. Indeed, if
this relation is true up to order $k-1$, then it is also true at order
$k$ because in
\begin{equation}
\delta \om{k} \sim [\om{r},\om{s}]_{\mbox{orig}} + [\om{r'},
\om{s'}]_{\ghp, \eta}
\end{equation}
($r+s = k-1, \; r' + s' = k$), the right hand side is of degree $\leq
-k+2$. Thus $\deg \om{k} \leq -k + 2 - 1 = -k + 1$.

But the element with most negative degree in the algebra is given by
the product of all the $\eta$'s, which has degree $-\sum_a(n_a - 1)$.
Accordingly, $\om{k}$ is zero whenever $-k+1 >= - \sum_a(n_a - 1)$,
which implies $r \leq \sum_a(n_a-1)+1$ as stated in the theorem.
\end{proof}

\noindent{\bf Remarks:}
\begin{enumerate}
\item One can improve greatly the bound by observing that the $\eta$'s
do not come alone in $\om{k}$. There are also $k$ momenta $\ghp_a$
which carry positive degree. This remark will, however, not be pursued
further here.
\item One can actually assign degrees smaller than one to the
generators $G_a$. For instance, in the case of an Abelian Lie algebra,
one may take $deg G_a = 1/2, \; \deg \eta^a = 1/2, \deg \ghp_a = -
1/2$. Because the degree of a ghost number one object is necessarily
greater than or equal to $1/2$, the condition $\deg \om{k} \leq -k+1$
(if $\om{k} \neq 0$) implies $\om{k} = 0$ for $k > 0$.
\end{enumerate}

\section{Self-reproducing algebras}

While Lie algebras are characterized by the existence of a degreee
that is decreased by the bracket, one may easily construct examples
of Poisson algebras for which such a degree does not exist. The
simplest ones are quadratic algebras for which $[G_a, G_b]$ is
proportional to $G_a, G_b$
\begin{equation}
[G_a, G_b] = M_{ab} G_a G_b \quad\quad\mbox{no summation on $a,b$}
\end{equation}
with $M_{ab} = -M_{ba}$. The Jacobi identity is fulfilled
for arbitrary $M$'s. Since $\deg(G_a G_b) = n_a + n_b$, the
inequality (\ref{deg-g=n}) is violated for any choice of
$n_a$. Because $[G_a, G_b]$ is proportional to $G_a G_b$, we shall
call these algebras ``self-reproducing algebras''.

The most general self-reproducing algebra with three generators is
given by
\begin{eqnarray}
[G_1, G_2] &=& \alpha\, G_1 G_2 \\{}
[G_2, G_3] &=& \beta \, G_2 G_3 \\{}
[G_3, G_1] &=& \gamma \, G_1 G_3.
\end{eqnarray}
This Poisson algebra can be realized on a six-dimensional phase space
by setting
\begin{equation}
G_1 = \exp(p_2 + \alpha q_3), G_2 = \exp(p_3 + \beta q_1), G_3 =
\exp(p_1 + \gamma q_2).
\end{equation}
The covariant BRST charge for this model is equal to
\begin{eqnarray}
\Omega &=&\eta^1 \, G_1 + \eta^2 \, G_2 + \eta^3 \, G_3 + \\
\nonumber
& & 1/2 \,
  (\alpha \,\eta^{2}\,\eta^{1}\,\ghp_{2}\,G_{1}
  -\alpha \,\eta^{2}\,\eta^{1}\,\ghp_{1}\,G_{2}
  -\beta \,\eta^{3}\,\eta^{2}\,\ghp_{3}\,G_{2} \\ \nonumber
& &\mbox{~~~~~~}
  -\beta \,\eta^{3}\,\eta^{2}\,\ghp_{2}\,G_{3}
  +\gamma \,\eta^{3}\,\eta^{1}\,\ghp_{3}\,G_{1}
  +\gamma \,\eta^{3}\,\eta^{1}\,\ghp_{1}\,G_{3}) + \\ \nonumber
& & 1/12 \, (
  ( - \alpha \,\beta +2\,\alpha \,\gamma  -\beta \,\gamma )
     \,\eta^{3}\,\eta^{2}\,\eta^{1}\,\ghp_{3}\,\ghp_{2}\,G_{1} + \\
\nonumber & &\mbox{~~~~~~~~~}
  ( -2\,\alpha \,\beta +\alpha \,\gamma +\beta \,\gamma )
     \,\eta^{3}\,\eta^{2}\,\eta^{1}\,\ghp_{3}\,\ghp_{1}\,G_{2} +\\
\nonumber & &\mbox{~~~~~~~~~}
  ( -\alpha \,\beta -\alpha \,\gamma +2\,\beta \,\gamma )
     \,\eta^{3}\,\eta^{2}\,\eta^{1}\,\ghp_{2}\,\ghp_{1}\,G_{3}
)
\end{eqnarray}
and is of rank 2 (the maximum possible rank) unless $\alpha = \beta =
\gamma$, or $\alpha = \beta = 0$, $\gamma \neq 0$, in which case it is
of rank 1.

\section{Examples}
We now give the BRST charge (or the first terms of the BRST charge)
for some particular Poisson algebras.  The examples have been treated
using REDUCE, using the treatment of summation over dummy indices
developed in \cite{Dre:CanExp,Dre:Imacs}. Details of the
implementation of the BRST algorithm can be found in \cite{BurCapDre:}.
All dummy variables are noted as $d_i$ where $i$ is an integer. Unless
stated otherwise, there is an implicit summation on all dummy
variables. For the examples in which the Jacobi identity is not
trivially satisfied,
the expressions have been normalized so that no
combinations of terms in a polynomial belongs to the polynomial ideal
generated by the left hand side of the Jacobi identity. In
particular, polynomials in this ideal are represented by identically null
expressions.

\subsection{Self-Reproducing Algebras}

As we have just defined, the basic Poisson brackets for the generators
$G_d$ of the {\em self-reproducing algebra} are given by
\begin{equation}
[G_{d_1}, G_{d_2}] = M_{d_1 d_2} G_{d_1} G_{d_2}
\end{equation}
without summation over the dummy variables $d_1$ and $d_2$. The matrix
$M$ is antisymmetric, but otherwise arbitrary.

The seven first orders of the covariant BRST charge are given by

$$\displaylines{\qdd
\om{0}=
\[G_{d_{1}}\,\eta^{d_{1}}
\]
\cr}$$
$$\displaylines{\qdd
\om{1}=
\[\frac{G_{d_{1}}\,M_{d_{1}d_{2}}\,\eta^{d_{1}}\,
        \eta^{d_{2}}\,\ghp_{d_{2}}}{
        2}
\]
\cr}$$
$$\displaylines{\qdd
\om{2}=
\[\frac{-
        \(G_{d_{1}}\,M_{d_{1}d_{2}}\,\eta^{d_{1}}\,
          \eta^{d_{2}}\,\eta^{d_{3}}\,\ghp_{d_{2}}\,
          \ghp_{d_{3}}\,
          \(M_{d_{1}d_{3}}
            +M_{d_{2}d_{3}}
          \)
        \)
        }{
        12}
\]
\cr}$$
$$\displaylines{\qdd
\om{3}=
\[\frac{-
        \(G_{d_{1}}\,M_{d_{1}d_{2}}\,M_{d_{1}d_{4}}\,
          M_{d_{2}d_{3}}\,\eta^{d_{1}}\,\eta^{d_{2}}\,
          \eta^{d_{3}}\,\eta^{d_{4}}\,\ghp_{d_{2}}\,
          \ghp_{d_{3}}\,\ghp_{d_{4}}
        \)
        }{
        24}
\]
\cr}$$
$$\displaylines{\qdd
\om{4}=
\[\(G_{d_{1}}\,\eta^{d_{1}}\,\eta^{d_{2}}\,
    \eta^{d_{3}}\,\eta^{d_{4}}\,\eta^{d_{5}}\,
    \ghp_{d_{2}}\,\ghp_{d_{3}}\,\ghp_{d_{4}}\,
    \ghp_{d_{5}}\,\nl
    \off{3499956}
    \(-
      \(M_{d_{1}d_{2}}\,M_{d_{1}d_{3}}\,M_{d_{1}d_{4}}\,
        M_{d_{1}d_{5}}
      \)
      +4\,M_{d_{1}d_{2}}\,M_{d_{1}d_{4}}\,M_{d_{1}d_{5}}\,
      M_{d_{2}d_{3}}\nl
      \off{3827636}
      +2\,M_{d_{1}d_{2}}\,M_{d_{1}d_{4}}\,M_{d_{2}d_{3}}\,
      M_{d_{4}d_{5}}
      +M_{d_{1}d_{2}}\,M_{d_{1}d_{5}}\,M_{d_{2}d_{3}}\nl
      \off{3827636}
      \,M_{d_{2}d_{4}}
      -M_{d_{1}d_{2}}\,M_{d_{2}d_{3}}\,M_{d_{2}d_{4}}\,
      M_{d_{2}d_{5}}
      -M_{d_{1}d_{4}}\,M_{d_{1}d_{5}}\,\nl
      \off{3827636}
      M_{d_{2}d_{3}}\,M_{d_{2}d_{4}}
      -2\,M_{d_{1}d_{4}}\,M_{d_{2}d_{3}}\,M_{d_{2}d_{4}}\,
      M_{d_{4}d_{5}}
      +M_{d_{1}d_{5}}\,\nl
      \off{3827636}
      M_{d_{2}d_{3}}\,M_{d_{2}d_{4}}\,M_{d_{2}d_{5}}
      -M_{d_{1}d_{5}}\,M_{d_{2}d_{3}}\,M_{d_{2}d_{4}}\,
      M_{d_{4}d_{5}}
    \)
  \)
  /720
\]
\Nl}$$
$$\displaylines{\qdd
\om{5}=
\[\(G_{d_{2}}\,M_{d_{1}d_{2}}\,M_{d_{3}d_{4}}\,
    \eta^{d_{1}}\,\eta^{d_{2}}\,\eta^{d_{3}}\,
    \eta^{d_{4}}\,\eta^{d_{5}}\,\eta^{d_{6}}\,
    \ghp_{d_{1}}\,\ghp_{d_{3}}\,\ghp_{d_{4}}\,
    \ghp_{d_{5}}\,\ghp_{d_{6}}\nl
    \off{3499956}
    \,
    \(-
      \(M_{d_{2}d_{3}}\,M_{d_{2}d_{5}}\,M_{d_{2}d_{6}}
      \)
      +2\,M_{d_{2}d_{3}}\,M_{d_{2}d_{5}}\,M_{d_{5}d_{6}}
      +M_{d_{2}d_{3}}\nl
      \off{4100703}
      \,M_{d_{2}d_{6}}\,M_{d_{3}d_{5}}
      -M_{d_{2}d_{3}}\,M_{d_{3}d_{5}}\,M_{d_{3}d_{6}}
      -M_{d_{2}d_{5}}\nl
      \off{4100703}
      \,M_{d_{2}d_{6}}\,M_{d_{3}d_{5}}
      -2\,M_{d_{2}d_{5}}\,M_{d_{3}d_{5}}\,M_{d_{5}d_{6}}\nl
      \off{4100703}
      +M_{d_{2}d_{6}}\,M_{d_{3}d_{5}}\,M_{d_{3}d_{6}}
      -M_{d_{2}d_{6}}\,M_{d_{3}d_{5}}\,M_{d_{5}d_{6}}
    \)
  \)
  /1440
\]
\Nl}$$
$$\displaylines{\qdd
\om{6}=
\[\(G_{d_{3}}\,\eta^{d_{1}}\,\eta^{d_{2}}\,
    \eta^{d_{3}}\,\eta^{d_{4}}\,\eta^{d_{5}}\,
    \eta^{d_{6}}\,\eta^{d_{7}}\,\ghp_{d_{1}}\,
    \ghp_{d_{2}}\,\ghp_{d_{4}}\,\ghp_{d_{5}}\,
    \ghp_{d_{6}}\,\ghp_{d_{7}}\,\nl
    \off{3499956}
    \(-
      \(M_{d_{1}d_{2}}\,M_{d_{1}d_{3}}\,M_{d_{1}d_{6}}\,
        M_{d_{2}d_{4}}\,M_{d_{4}d_{5}}\,M_{d_{6}d_{7}}
      \)
      -M_{d_{1}d_{2}}\,M_{d_{1}d_{3}}\nl
      \off{3827636}
      \,M_{d_{2}d_{4}}\,M_{d_{2}d_{6}}\,M_{d_{4}d_{5}}\,
      M_{d_{6}d_{7}}
      +2\,M_{d_{1}d_{2}}\,M_{d_{1}d_{3}}\,M_{d_{2}d_{4}}\,
      \nl
      \off{3827636}
      M_{d_{4}d_{5}}\,M_{d_{4}d_{6}}\,M_{d_{4}d_{7}}
      +M_{d_{1}d_{2}}\,M_{d_{1}d_{3}}\,M_{d_{2}d_{6}}\,
      M_{d_{4}d_{5}}\,M_{d_{4}d_{6}}\nl
      \off{3827636}
      \,M_{d_{6}d_{7}}
      -2\,M_{d_{1}d_{2}}\,M_{d_{1}d_{3}}\,M_{d_{2}d_{7}}\,
      M_{d_{4}d_{5}}\,M_{d_{4}d_{6}}\,M_{d_{4}d_{7}}\nl
      \off{3827636}
      +2\,M_{d_{1}d_{2}}\,M_{d_{1}d_{3}}\,M_{d_{2}d_{7}}\,
      M_{d_{4}d_{5}}\,M_{d_{4}d_{6}}\,M_{d_{6}d_{7}}\nl
      \off{3827636}
      -2\,M_{d_{1}d_{2}}\,M_{d_{2}d_{3}}\,M_{d_{2}d_{4}}\,
      M_{d_{2}d_{6}}\,M_{d_{4}d_{5}}\,M_{d_{6}d_{7}}\nl
      \off{3827636}
      -13\,M_{d_{1}d_{2}}\,M_{d_{2}d_{3}}\,
      M_{d_{2}d_{4}}\,M_{d_{3}d_{6}}\,M_{d_{4}d_{5}}\,
      M_{d_{6}d_{7}}\nl
      \off{3827636}
      +4\,M_{d_{1}d_{2}}\,M_{d_{2}d_{3}}\,M_{d_{2}d_{4}}\,
      M_{d_{4}d_{5}}\,M_{d_{4}d_{6}}\,M_{d_{4}d_{7}}\nl
      \off{3827636}
      +2\,M_{d_{1}d_{2}}\,M_{d_{2}d_{3}}\,M_{d_{2}d_{6}}\,
      M_{d_{4}d_{5}}\,M_{d_{4}d_{6}}\,M_{d_{6}d_{7}}\nl
      \off{3827636}
      -4\,M_{d_{1}d_{2}}\,M_{d_{2}d_{3}}\,M_{d_{2}d_{7}}\,
      M_{d_{4}d_{5}}\,M_{d_{4}d_{6}}\,M_{d_{4}d_{7}}\nl
      \off{3827636}
      +4\,M_{d_{1}d_{2}}\,M_{d_{2}d_{3}}\,M_{d_{2}d_{7}}\,
      M_{d_{4}d_{5}}\,M_{d_{4}d_{6}}\,M_{d_{6}d_{7}}\nl
      \off{3827636}
      -2\,M_{d_{1}d_{2}}\,M_{d_{2}d_{3}}\,M_{d_{3}d_{4}}\,
      M_{d_{3}d_{6}}\,M_{d_{4}d_{5}}\,M_{d_{6}d_{7}}\nl
      \off{3827636}
      +M_{d_{1}d_{3}}\,M_{d_{1}d_{4}}\,M_{d_{1}d_{6}}\,
      M_{d_{2}d_{6}}\,M_{d_{4}d_{5}}\,M_{d_{6}d_{7}}
      -2\,M_{d_{1}d_{3}}\nl
      \off{3827636}
      \,M_{d_{1}d_{4}}\,M_{d_{2}d_{4}}\,M_{d_{4}d_{5}}\,
      M_{d_{4}d_{6}}\,M_{d_{4}d_{7}}
      +5\,M_{d_{1}d_{3}}\,M_{d_{1}d_{4}}\,\nl
      \off{3827636}
      M_{d_{2}d_{7}}\,M_{d_{3}d_{7}}\,M_{d_{4}d_{5}}\,
      M_{d_{4}d_{6}}
      +M_{d_{1}d_{3}}\,M_{d_{1}d_{6}}\,M_{d_{2}d_{3}}\,
      M_{d_{2}d_{4}}\nl
      \off{3827636}
      \,M_{d_{4}d_{5}}\,M_{d_{6}d_{7}}
      -2\,M_{d_{1}d_{3}}\,M_{d_{1}d_{6}}\,M_{d_{2}d_{6}}\,
      M_{d_{4}d_{5}}\,M_{d_{4}d_{6}}\nl
      \off{3827636}
      \,M_{d_{6}d_{7}}
      -5\,M_{d_{1}d_{3}}\,M_{d_{1}d_{6}}\,M_{d_{2}d_{7}}\,
      M_{d_{3}d_{7}}\,M_{d_{4}d_{5}}\,M_{d_{4}d_{6}}\nl
      \off{3827636}
      +M_{d_{1}d_{3}}\,M_{d_{1}d_{7}}\,M_{d_{2}d_{7}}\,
      M_{d_{4}d_{5}}\,M_{d_{4}d_{6}}\,M_{d_{4}d_{7}}
      -M_{d_{1}d_{3}}\,M_{d_{1}d_{7}}\nl
      \off{3827636}
      \,M_{d_{2}d_{7}}\,M_{d_{4}d_{5}}\,M_{d_{4}d_{6}}\,
      M_{d_{6}d_{7}}
      -M_{d_{1}d_{3}}\,M_{d_{2}d_{3}}\,M_{d_{2}d_{4}}\,
      M_{d_{2}d_{6}}\,\nl
      \off{3827636}
      M_{d_{4}d_{5}}\,M_{d_{6}d_{7}}
      -8\,M_{d_{1}d_{3}}\,M_{d_{2}d_{3}}\,M_{d_{2}d_{4}}\,
      M_{d_{3}d_{6}}\,M_{d_{4}d_{5}}\,M_{d_{6}d_{7}}\nl
      \off{3827636}
      +2\,M_{d_{1}d_{3}}\,M_{d_{2}d_{3}}\,M_{d_{2}d_{4}}\,
      M_{d_{4}d_{5}}\,M_{d_{4}d_{6}}\,M_{d_{4}d_{7}}\nl
      \off{3827636}
      +M_{d_{1}d_{3}}\,M_{d_{2}d_{3}}\,M_{d_{2}d_{6}}\,
      M_{d_{4}d_{5}}\,M_{d_{4}d_{6}}\,M_{d_{6}d_{7}}
      -2\,M_{d_{1}d_{3}}\nl
      \off{3827636}
      \,M_{d_{2}d_{3}}\,M_{d_{2}d_{7}}\,M_{d_{4}d_{5}}\,
      M_{d_{4}d_{6}}\,M_{d_{4}d_{7}}
      +2\,M_{d_{1}d_{3}}\,M_{d_{2}d_{3}}\nl
      \off{3827636}
      \,M_{d_{2}d_{7}}\,M_{d_{4}d_{5}}\,M_{d_{4}d_{6}}\,
      M_{d_{6}d_{7}}
      -2\,M_{d_{1}d_{3}}\,M_{d_{2}d_{3}}\,M_{d_{3}d_{4}}\,
      \nl
      \off{3827636}
      M_{d_{3}d_{5}}\,M_{d_{3}d_{6}}\,M_{d_{3}d_{7}}
      +12\,M_{d_{1}d_{3}}\,M_{d_{2}d_{3}}\,
      M_{d_{3}d_{4}}\,M_{d_{3}d_{6}}\nl
      \off{3827636}
      \,M_{d_{3}d_{7}}\,M_{d_{4}d_{5}}
      -6\,M_{d_{1}d_{3}}\,M_{d_{2}d_{3}}\,M_{d_{3}d_{4}}\,
      M_{d_{3}d_{6}}\,M_{d_{4}d_{5}}\nl
      \off{3827636}
      \,M_{d_{6}d_{7}}
      -5\,M_{d_{1}d_{3}}\,M_{d_{2}d_{3}}\,M_{d_{3}d_{4}}\,
      M_{d_{3}d_{7}}\,M_{d_{4}d_{5}}\,M_{d_{4}d_{6}}\nl
      \off{3827636}
      +5\,M_{d_{1}d_{3}}\,M_{d_{2}d_{3}}\,M_{d_{3}d_{4}}\,
      M_{d_{4}d_{5}}\,M_{d_{4}d_{6}}\,M_{d_{4}d_{7}}\nl
      \off{3827636}
      +5\,M_{d_{1}d_{3}}\,M_{d_{2}d_{3}}\,M_{d_{3}d_{6}}\,
      M_{d_{3}d_{7}}\,M_{d_{4}d_{5}}\,M_{d_{4}d_{6}}\nl
      \off{3827636}
      +13\,M_{d_{1}d_{3}}\,M_{d_{2}d_{3}}\,
      M_{d_{3}d_{6}}\,M_{d_{4}d_{5}}\,M_{d_{4}d_{6}}\,
      M_{d_{6}d_{7}}\nl
      \off{3827636}
      -5\,M_{d_{1}d_{3}}\,M_{d_{2}d_{3}}\,M_{d_{3}d_{7}}\,
      M_{d_{4}d_{5}}\,M_{d_{4}d_{6}}\,M_{d_{4}d_{7}}\nl
      \off{3827636}
      +5\,M_{d_{1}d_{3}}\,M_{d_{2}d_{3}}\,M_{d_{3}d_{7}}\,
      M_{d_{4}d_{5}}\,M_{d_{4}d_{6}}\,M_{d_{6}d_{7}}\nl
      \off{3827636}
      +2\,M_{d_{1}d_{3}}\,M_{d_{2}d_{4}}\,M_{d_{3}d_{4}}\,
      M_{d_{4}d_{5}}\,M_{d_{4}d_{6}}\,M_{d_{4}d_{7}}\nl
      \off{3827636}
      -8\,M_{d_{1}d_{3}}\,M_{d_{2}d_{6}}\,M_{d_{3}d_{4}}\,
      M_{d_{3}d_{6}}\,M_{d_{4}d_{5}}\,M_{d_{6}d_{7}}\nl
      \off{3827636}
      +2\,M_{d_{1}d_{3}}\,M_{d_{2}d_{6}}\,M_{d_{3}d_{6}}\,
      M_{d_{4}d_{5}}\,M_{d_{4}d_{6}}\,M_{d_{6}d_{7}}\nl
      \off{3827636}
      -M_{d_{1}d_{3}}\,M_{d_{2}d_{7}}\,M_{d_{3}d_{7}}\,
      M_{d_{4}d_{5}}\,M_{d_{4}d_{6}}\,M_{d_{4}d_{7}}\nl
      \off{3827636}
      +M_{d_{1}d_{3}}\,M_{d_{2}d_{7}}\,M_{d_{3}d_{7}}\,
      M_{d_{4}d_{5}}\,M_{d_{4}d_{6}}\,M_{d_{6}d_{7}}
      -2\,M_{d_{1}d_{4}}\,\nl
      \off{3827636}
      M_{d_{2}d_{4}}\,M_{d_{3}d_{4}}\,M_{d_{4}d_{5}}\,
      M_{d_{4}d_{6}}\,M_{d_{4}d_{7}}
      +M_{d_{1}d_{4}}\,M_{d_{2}d_{6}}\,M_{d_{3}d_{4}}\nl
      \off{3827636}
      \,M_{d_{3}d_{6}}\,M_{d_{4}d_{5}}\,M_{d_{6}d_{7}}
      +5\,M_{d_{1}d_{6}}\,M_{d_{2}d_{6}}\,M_{d_{3}d_{4}}\,
      M_{d_{3}d_{6}}\nl
      \off{3827636}
      \,M_{d_{4}d_{5}}\,M_{d_{6}d_{7}}
      -8\,M_{d_{1}d_{6}}\,M_{d_{2}d_{6}}\,M_{d_{3}d_{6}}\,
      M_{d_{4}d_{5}}\,M_{d_{4}d_{6}}\nl
      \off{3827636}
      \,M_{d_{6}d_{7}}
      -2\,M_{d_{1}d_{7}}\,M_{d_{2}d_{7}}\,M_{d_{3}d_{6}}\,
      M_{d_{3}d_{7}}\,M_{d_{4}d_{5}}\,M_{d_{4}d_{6}}\nl
      \off{3827636}
      +6\,M_{d_{1}d_{7}}\,M_{d_{2}d_{7}}\,M_{d_{3}d_{7}}\,
      M_{d_{4}d_{5}}\,M_{d_{4}d_{6}}\,M_{d_{4}d_{7}}
      -6\,\nl
      \off{3827636}
      M_{d_{1}d_{7}}\,M_{d_{2}d_{7}}\,M_{d_{3}d_{7}}\,
      M_{d_{4}d_{5}}\,M_{d_{4}d_{6}}\,M_{d_{6}d_{7}}
    \)
  \)
  /60480
\]
\Nl}$$

These expressions are not particularly illuminating but are of interest
because they generically do not vanish and hence, define higher
order BRST charges. This can be seen by means
of the following example, in which only the brackets of the first
generator with the other ones are non vanishing, and taken equal to
\begin{eqnarray}\label{eq:maxquad}
[G_1, G_{\alpha}] &=& G_1 G_{\alpha} = - [ G_{\alpha}, G_1]
    \quad  (\alpha = 2,3,\ldots,n),\\{}
[G_{\alpha}, G_{\beta}] &=& 0.
\end{eqnarray}
For this particular self-reproducing algebra, all orders of the
covariant BRST charge can be explicitly computed. One finds
\begin{eqnarray}
\om{0} &=& \eta^a G_a \\
\om{k} &=& \alpha_k (\ov{T_1}{(k)} + (-)^{k+1} \ov{T_2}{(k)})
\end{eqnarray}
where
\begin{eqnarray}
\ov{T_1}{(k)} &=& G_1 \eta^{\alpha_1} \cdots \eta^{\alpha_k} \eta^1
                 \ghp_{\alpha_1} \cdots \ghp_{\alpha_k} \\
\ov{T_2}{(k)} &=& G_{\alpha_k} \eta^{\alpha_1} \cdots \eta^{\alpha_k} \eta^1
                 \ghp_1 \ghp_{\alpha_1} \cdots \ghp_{\alpha_{k-1}}
\end{eqnarray}
and
\begin{eqnarray}
\alpha_1 &=& -1/2, \; \alpha_2 = -1/12, \; \alpha_3 = 0 \\
\alpha_k &=& -\frac{1}{k+1} \sum_{l=1}^{k-3} \alpha_{l+1} \alpha_{k-l-1}
            \quad \mbox{for $k > 3$}\label{eq:AlpRec}.
\end{eqnarray}

This can be seen as the only non zero brackets involved in the
construction of the BRST charge are
\begin{eqnarray}
[\ov{T_1}{(k)}, \om{0}]_{\mbox{orig}} &=& (-)^{k+1}\ov{S}{(k)} \\{}
[\ov{T_1}{(k)}, \ov{T_2}{(l)}]_{\eta\ghp} &=&
  (-)^{l(k+1)}\ov{S}{(k+l-1)},
\end{eqnarray}
where
\begin{equation}
\ov{S}{(k)} = G_1 G_\alpha \eta^{\alpha_1} \cdots \eta^{\alpha_k} \eta^\alpha
\eta^1
                 \ghp_{\alpha_1} \cdots \ghp{\alpha_k}.
\end{equation}
We further have
\begin{equation}\label{eq:sigmaS}
\sigma \ov{S}{(k)} = (-)^{k+1} \ov{T_2}{(k+1)} - \ov{T_1}{(k+1)}
\end{equation}
Given these relations, it is straightforward to verify (39).  First, one
easily checks that (39) is correct for $k=1$
 with $\alpha_1$ equal to 1/2. Let us then
assume that (39) is true for $k = 0, 1 ...$ up to $p$.  One then obtains
\begin{equation}
\ov{D}{(p)} = \beta_p  \ov{S}{(p)}
\end{equation}
with $\beta_p$ given by
\begin{equation}
\beta_p = (-)^{p+1} \alpha_p - \sum_{k=0}^{p-1} (-)^{p(k+1)} \alpha_{k+1}
\alpha_{p-k} = - \sum_{k=1}^{p-2} (-)^{p(k+1)} \alpha_{k+1} \alpha_{p-k}
\end{equation}
from which one gets, using (\ref{eq:sigmaS}), that $\om{p+1}$ is indeed given
by (39) with $\alpha_{p+1}$ equal to
\begin{equation}
\alpha_{p+1} = \frac{\beta_p}{p+2}
\end{equation}
Observe now that  $\alpha_k = 0$ for $k$ odd, $k \neq 1$. This can again be
shown by recurrence.  First note that $\alpha_3$ = 0.  Now let $p$ be
even, $p > 3$. Suppose $\alpha_k = 0$ for $k$ odd, $1 < k < p$. All
terms in the relation defining $\beta_{p}$ are proportional to an
$\alpha_m$ with $m$ odd, $1 < m < p$, since $k+1$ and $p-k$ have
opposite parities.  Therefore, $\beta_p = 0 = \alpha_{p+1}$ and thus $\alpha_k
= 0$ for $k$ odd, $k > 1$.
Accordingly only $\alpha_k $ with $k$ even can be different from zero.
The expression for $\alpha_k$ reduces then to (43) since $k+1$ must
be even in (49).

Although $\alpha_k = 0$ for $k$ odd, $k > 1$, one easily sees that
 $\alpha_k < 0$ for $k$ even. This is true for $k =
2$ as $\alpha_2 = -1/12$. Let $p$ be even, and suppose $\alpha_m < 0$
for $1 < m < p$, $m$ even. Then, all terms in the sum in the
recurrence relation (\ref{eq:AlpRec}) are strictly positive, so that
$\alpha_p < 0$. Since $\alpha_p \neq 0$, the quadratic
algebra (\ref{eq:maxquad}) provides examples of systems with
arbitrarily high covariant rank.

Note also that the minimal rank is equal to one: indeed, the
non covariant BRST charge given by
\begin{equation}
\tilde{\Omega}=\om{0} + \om{1} +  (\TT{1}_1 - \TT{1}_2)/2 =
\eta^a G_a - G_a \eta^a \eta^1 \ghp_1
\end{equation}
is nilpotent and
\begin{equation}
\delta(\TT{1}_1 - \TT{1}_2) = 0
\end{equation}
so $\tilde{\Omega}$ is indeed a valid BRST charge.  This shows that the minimal
and covariant ranks are in general different.

Finally, it is easy to modify slightly the basic brackets so as to
induce non zero covariant $\om{k}$ with $k$ odd.  One simply replaces
(\ref{eq:maxquad}) by
\begin{eqnarray}
[G_{n-1}, G_n] &=& G_{n-1} G_n = - [G_n, G_{n-1}] \\{}
[G_1, G_n] &=& - G_1 G_n, \\{}
[G_1, G_{\alpha}] &=& G_1 G_{\alpha} \quad (\alpha \neq n).
\end{eqnarray}

\subsection{Purely Quadratic Algebras}
A generalization of the above is the pure quadratic algebra. The basic
Poisson brackets are then given by
\begin{equation}
[G_{d_1}, G_{d_2}] = D_{d_1 d_2}^{d_3 d_4} G_{d_3} G_{d_4}
\end{equation}
where $D_{d_1 d_2}^{d_3 d_4}$ is antisymmetric in $d_1, d_2$ and
symmetric in $d_3, d_4$. The Jacobi identity implies that
\begin{equation}
D_{d_4 d_1}^{d_5 d_6} D_{d_2 d_3}^{d_4 d_7} + \mbox{symm}(d_5, d_6,
d_7) + \mbox{cyclic}(d_1,d_2,d_3) = 0
\end{equation}

The first orders of the covariant BRST charge are given by

$$\displaylines{\qdd
\om{0}=
\[G_{d_{1}}\,\eta^{d_{1}}
\]
\cr}$$
$$\displaylines{\qdd
\om{1}=
\[\frac{D_{d_{1}d_{2}}^{
           d_{4}d_{3}}\,G_{d_{4}}\,\eta^{d_{1}}\,
        \eta^{d_{2}}\,\ghp_{d_{3}}}{
        2}
\]
\cr}$$
$$\displaylines{\qdd
\om{2}=
\[\frac{D_{d_{2}d_{3}}^{
           d_{6}d_{5}}\,D_{d_{6}d_{1}}^{
                           d_{7}d_{4}}\,G_{d_{7}}\,
        \eta^{d_{1}}\,\eta^{d_{2}}\,\eta^{d_{3}}\,
        \ghp_{d_{4}}\,\ghp_{d_{5}}}{
        6}
\]
\cr}$$
$$\displaylines{\qdd
\om{3}=
\[\frac{-
        \(D_{d_{2}d_{3}}^{
             d_{8}d_{7}}\,D_{d_{4}d_{1}}^{
                             d_{9}d_{5}}\,D_{d_{8}d_{9}}^{
          d_{10}d_{6}}\,G_{d_{10}}\,\eta^{d_{1}}\,
          \eta^{d_{2}}\,\eta^{d_{3}}\,\eta^{d_{4}}\,
          \ghp_{d_{5}}\,\ghp_{d_{6}}\,\ghp_{d_{7}}
        \)
        }{
        24}
\]
\cr}$$
$$\displaylines{\qdd
\om{4}=
\[\(\eta^{d_{1}}\,\eta^{d_{2}}\,\eta^{d_{3}}\,
    \eta^{d_{4}}\,\eta^{d_{5}}\,\ghp_{d_{6}}\,
    \ghp_{d_{7}}\,\ghp_{d_{8}}\,\ghp_{d_{9}}\,
    \nl
    \off{3499956}
    \(3\,D_{d_{1}d_{2}}^{
            d_{13}d_{6}}\,D_{d_{4}d_{5}}^{
                             d_{12}d_{9}}\,
      D_{d_{10}d_{3}}^{
         d_{11}d_{8}}\,D_{d_{12}d_{13}}^{
                          d_{10}d_{7}}\,G_{d_{11}}
      +4\,D_{d_{3}d_{4}}^{
             d_{10}d_{9}}\,D_{d_{5}d_{2}}^{
                              d_{13}d_{6}}\,
      D_{d_{10}d_{11}}^{
         d_{12}d_{8}}\nl
      \off{3827636}
      \,D_{d_{13}d_{1}}^{
           d_{11}d_{7}}\,G_{d_{12}}
      -4\,D_{d_{4}d_{5}}^{
             d_{12}d_{9}}\,D_{d_{10}d_{3}}^{
                              d_{11}d_{8}}\,
      D_{d_{12}d_{1}}^{
         d_{13}d_{7}}\,D_{d_{13}d_{2}}^{
                          d_{10}d_{6}}\,G_{d_{11}}
    \)
  \)
  /360
\]
\Nl}$$
$$\displaylines{\qdd
\om{5}=
\[\(D_{d_{12}d_{13}}^{
       d_{14}d_{10}}\,G_{d_{14}}\,\eta^{d_{1}}\,
    \eta^{d_{2}}\,\eta^{d_{3}}\,\eta^{d_{4}}\,
    \eta^{d_{5}}\,\eta^{d_{6}}\,\ghp_{d_{7}}\,
    \ghp_{d_{8}}\,\ghp_{d_{9}}\,\ghp_{d_{10}}\,
    \ghp_{d_{11}}\nl
    \off{3499956}
    \,
    \(-
      \(2\,D_{d_{1}d_{2}}^{
              d_{16}d_{7}}\,D_{d_{5}d_{6}}^{
                               d_{15}d_{11}}\,
        D_{d_{15}d_{3}}^{
           d_{12}d_{8}}\,D_{d_{16}d_{4}}^{
                            d_{13}d_{9}}
      \)
      +3\,D_{d_{1}d_{3}}^{
             d_{16}d_{8}}\,D_{d_{4}d_{5}}^{
                              d_{12}d_{11}}\,
      \nl
      \off{4100703}
      D_{d_{6}d_{2}}^{
         d_{15}d_{7}}\,D_{d_{15}d_{16}}^{
                          d_{13}d_{9}}
      -4\,D_{d_{4}d_{5}}^{
             d_{12}d_{11}}\,D_{d_{6}d_{1}}^{
                               d_{15}d_{7}}\,
      D_{d_{15}d_{3}}^{
         d_{16}d_{9}}\,D_{d_{16}d_{2}}^{
                          d_{13}d_{8}}
    \)
  \)
  /720
\]
\Nl}$$
$$\displaylines{\qdd
\om{6}=
\[\(\eta^{d_{1}}\,\eta^{d_{2}}\,\eta^{d_{3}}\,
    \eta^{d_{4}}\,\eta^{d_{5}}\,\eta^{d_{6}}\,
    \eta^{d_{7}}\,\ghp_{d_{8}}\,\ghp_{d_{9}}\,
    \ghp_{d_{10}}\,\ghp_{d_{11}}\,\ghp_{d_{12}}\,
    \ghp_{d_{13}}\,\nl
    \off{3499956}
    \(6\,D_{d_{1}d_{2}}^{
            d_{17}d_{8}}\,D_{d_{6}d_{7}}^{
                             d_{16}d_{13}}\,
      D_{d_{14}d_{5}}^{
         d_{15}d_{12}}\,D_{d_{16}d_{17}}^{
                           d_{18}d_{11}}\,D_{d_{18}d_{4}}^{
      d_{19}d_{10}}\,D_{d_{19}d_{3}}^{
                        d_{14}d_{9}}\,G_{d_{15}}\nl
      \off{3827636}
      -9\,D_{d_{1}d_{2}}^{
             d_{18}d_{8}}\,D_{d_{3}d_{4}}^{
                              d_{19}d_{9}}\,
      D_{d_{6}d_{7}}^{
         d_{16}d_{13}}\,D_{d_{14}d_{5}}^{
                           d_{15}d_{12}}\,D_{d_{16}d_{17}}^{
      d_{14}d_{10}}\,D_{d_{18}d_{19}}^{
                        d_{17}d_{11}}\,G_{d_{15}}\nl
      \off{3827636}
      +12\,D_{d_{1}d_{2}}^{
              d_{18}d_{8}}\,D_{d_{3}d_{4}}^{
                               d_{19}d_{9}}\,
      D_{d_{6}d_{7}}^{
         d_{17}d_{13}}\,D_{d_{14}d_{15}}^{
                           d_{16}d_{12}}\,D_{d_{17}d_{18}}^{
      d_{15}d_{11}}\,D_{d_{19}d_{5}}^{
                        d_{14}d_{10}}\,G_{d_{16}}\nl
      \off{3827636}
      -10\,D_{d_{1}d_{2}}^{
              d_{18}d_{8}}\,D_{d_{6}d_{7}}^{
                               d_{16}d_{13}}\,
      D_{d_{14}d_{5}}^{
         d_{15}d_{12}}\,D_{d_{16}d_{3}}^{
                           d_{17}d_{9}}\,D_{d_{17}d_{19}}^{
      d_{14}d_{11}}\,D_{d_{18}d_{4}}^{
                        d_{19}d_{10}}\,G_{d_{15}}\nl
      \off{3827636}
      +24\,D_{d_{1}d_{2}}^{
              d_{18}d_{8}}\,D_{d_{6}d_{7}}^{
                               d_{16}d_{13}}\,
      D_{d_{14}d_{5}}^{
         d_{15}d_{12}}\,D_{d_{16}d_{17}}^{
                           d_{14}d_{11}}\,D_{d_{18}d_{4}}^{
      d_{19}d_{10}}\,D_{d_{19}d_{3}}^{
                        d_{17}d_{9}}\,G_{d_{15}}\nl
      \off{3827636}
      -4\,D_{d_{1}d_{2}}^{
             d_{18}d_{8}}\,D_{d_{6}d_{7}}^{
                              d_{17}d_{13}}\,
      D_{d_{14}d_{15}}^{
         d_{16}d_{12}}\,D_{d_{17}d_{4}}^{
                           d_{15}d_{9}}\,D_{d_{18}d_{5}}^{
      d_{19}d_{11}}\,D_{d_{19}d_{3}}^{
                        d_{14}d_{10}}\,G_{d_{16}}\nl
      \off{3827636}
      +4\,D_{d_{1}d_{2}}^{
             d_{19}d_{8}}\,D_{d_{6}d_{7}}^{
                              d_{16}d_{13}}\,
      D_{d_{14}d_{5}}^{
         d_{15}d_{12}}\,D_{d_{16}d_{17}}^{
                           d_{18}d_{11}}\,D_{d_{18}d_{3}}^{
      d_{14}d_{9}}\,D_{d_{19}d_{4}}^{
                       d_{17}d_{10}}\nl
      \off{3827636}
      \,G_{d_{15}}
      -6\,D_{d_{1}d_{3}}^{
             d_{18}d_{9}}\,D_{d_{5}d_{6}}^{
                              d_{14}d_{13}}\,
      D_{d_{7}d_{2}}^{
         d_{17}d_{8}}\,D_{d_{14}d_{15}}^{
                          d_{16}d_{12}}\,D_{d_{17}d_{18}}^{
      d_{19}d_{11}}\nl
      \off{3827636}
      \,D_{d_{19}d_{4}}^{
           d_{15}d_{10}}\,G_{d_{16}}
      -4\,D_{d_{1}d_{3}}^{
             d_{19}d_{9}}\,D_{d_{5}d_{6}}^{
                              d_{14}d_{13}}\,
      D_{d_{7}d_{2}}^{
         d_{17}d_{8}}\,D_{d_{14}d_{15}}^{
                          d_{16}d_{12}}\,\nl
      \off{3827636}
      D_{d_{17}d_{18}}^{
         d_{15}d_{11}}\,D_{d_{19}d_{4}}^{
                           d_{18}d_{10}}\,G_{d_{16}}
      -16\,D_{d_{5}d_{6}}^{
              d_{14}d_{13}}\,D_{d_{7}d_{1}}^{
                                d_{17}d_{8}}\,
      D_{d_{14}d_{15}}^{
         d_{16}d_{12}}\nl
      \off{3827636}
      \,D_{d_{17}d_{4}}^{
           d_{18}d_{11}}\,D_{d_{18}d_{3}}^{
                             d_{19}d_{10}}\,
      D_{d_{19}d_{2}}^{
         d_{15}d_{9}}\,G_{d_{16}}
      +16\,D_{d_{6}d_{7}}^{
              d_{16}d_{13}}\,D_{d_{14}d_{5}}^{
                                d_{15}d_{12}}\nl
      \off{3827636}
      \,D_{d_{16}d_{4}}^{
           d_{17}d_{11}}\,D_{d_{17}d_{3}}^{
                             d_{19}d_{10}}\,
      D_{d_{18}d_{1}}^{
         d_{14}d_{8}}\,D_{d_{19}d_{2}}^{
                          d_{18}d_{9}}\,G_{d_{15}}
    \)
  \)
  /15120
\]
\Nl}$$

Again, these expressions are not particularly illuminating.  The point
emphasized here is that the calculation of the BRST charge is purely
algorithmic and follows a general, well-established pattern.

Since homogeneous quadratic
 algebras contain the self-reproducing algebras as special
case, they are generically of maximal covariant rank. More on this
in \cite{ADresse3}.

\subsection{L-T algebras}
We now consider adding a linear term to the quadratic algebra
above. The basic Poisson brackets for the generators $G_d$ are given
by
\begin{equation}
[G_{d_1}, G_{d_2}] = C_{d_1 d_2}^{d_3}  \, G_{d_3} +
   D_{d_1 d_2}^{d_3 d_4} G_{d_3} G_{d_4}
\end{equation}
where $ C_{d_1 d_2}^{d_3} $ and $D_{d_1 d_2}^{d_3 d_4}$ are
antisymmetric in $d_1, d_2$, and $D_{d_1 d_2}^{d_3 d_4}$ is symmetric
in $d_3, d_4$.  A particular instance of such an algebra is given
by Zamolodchikov algebras \cite{Zam:}.  We will start with a specific
example, and consider general quadratically nonlinear Poisson algebras
next.

The generators in the example are assumed to split into $L_{a}$ and
$T_b$, $a = 1, \ldots n_1$, $b = n_1 + 1, \ldots, n$, with the
brackets
\begin{eqnarray}
[L_{a_1}, L_{a_2}] &=& \tilde{C}_{a_1 a_2}^{a_3} L_{a_3} \nonumber \\{}
[L_{a_1}, T_{b_1}] &=& \tilde{C}_{a_1 b_1}^{a_2} L_{a_2} +
                       \tilde{C}_{a_1 b_1}^{b_2} T_{b_2}
  \label{eq:zam} \\{}
[T_{b_1}, T_{b_2}] &=& \tilde{C}_{b_1 b_2}^{a_1} L_{a_1} +
                       \tilde{C}_{b_1 b_2}^{b_3} T_{b_3} +
                       \tilde{D}_{b_1  b_2}^{a_1 a_2} L_{a_1} L_{a_2}
\nonumber
\end{eqnarray}
so that contractions of $\tilde{D}$ are impossible.

Going back to the notations $G_{d_i} = \{L_{a}, T_{b}\}$, $d = 1,
\ldots, n$ the Jacobi identity imply
\begin{equation}
C_{d_1 d_2}^{d_4} C_{d_3 d_4}^{d_5} + \mbox{cyclic}(d_1,d_2,d_3) = 0
\end{equation}
\begin{equation}
\{D_{d_1 d_2}^{d_4 d_5} C_{d_3 d_4}^{d_6} + \mbox{symm}(d_5, d_6) \} + \\
  C_{d_1 d_2}^{d_4} D_{d_3 d_4}^{d_5 d_6} + \mbox{cyclic}(d_1,d_2,d_3) = 0
\end{equation}
and contractions of $D$ vanish.

For instance, the conditions (\ref{eq:zam}) are fulfilled if one takes
for the $L$'s the generators of a semi-simple Lie algebra and take the
$T$'s to commute with the $L$'s and to close on the Casimir element:
\begin{eqnarray}
[L_a, T_b] &=& 0 \\{}
[T_{b_1}, T_{b_2}] &=& \delta_{b_1 b_2} k^{a_1 a_2} L_{a_1} L_{a_2}
\end{eqnarray}
where $k^{a_1 a_2}$ is the Killing bilinear form. The Jacobi identity
is verified because the Casimir element commutes with the $L$'s.

The previous theorem on
the rank yields, by taking $n(l) = 1$ and $n(T) = 3/2$, that the rank
is bounded by $1/2 m + 1$, where $m$ is the number of T-generators.
Actually, the rank is much lower, since the covariant BRST charge is
computed to be
$$\displaylines{\qdd
\Omega =
\[\frac{1}{2}
  \,C_{d_{1}d_{2}}^{
       d_{3}}\,\eta^{d_{1}}\,\eta^{d_{2}}\,
  \ghp_{d_{3}}
  +
  \frac{1}{24}
  \,C_{d_{8}d_{9}}^{
       d_{6}}\,D_{d_{1}d_{2}}^{
                  d_{8}d_{7}}\,D_{d_{3}d_{4}}^{
  d_{9}d_{5}}\,\eta^{d_{1}}\,\eta^{d_{2}}\nl
  \off{2695321}
  \,\eta^{d_{3}}\,\eta^{d_{4}}\,\ghp_{d_{5}}\,
  \ghp_{d_{7}}\,\ghp_{d_{6}}
  +
  \frac{1}{2}
  \,D_{d_{1}d_{2}}^{
       d_{4}d_{3}}\,G_{d_{4}}\,\eta^{d_{1}}\,
  \eta^{d_{2}}\,\ghp_{d_{3}}
  +G_{d_{1}}\,\eta^{d_{1}}
\]
\Nl}$$
which is identical to the result in \cite{SchSevNie:QuaBRSChaQua}.

\subsection{Generalizations}
The previous L-T algebras can be generalized in various directions.
One may consider the general quadratic non homogeneous Poisson
structure
\begin{eqnarray}
&  [G_{d_1}, G_{d_2}] = C_{d_1 d_2}^{d_3}  \, G_{d_3} +
     D_{d_1 d_2}^{d_3 d_4} G_{d_3} G_{d_4} & \\
&  C_{d_1 d_2}^{d_4} C_{d_3 d_4}^{d_5} + \mbox{cyclic}(d_1,d_2,d_3)
    = 0 &  \\
&  \{D_{d_1 d_2}^{d_4 d_5} C_{d_3 d_4}^{d_6} + \mbox{symm}(d_5, d_6) \} +
    C_{d_1 d_2}^{d_4} D_{d_3 d_4}^{d_5 d_6} + \mbox{} \quad\quad &
    \nonumber \\
& \quad \quad \mbox{cyclic}(d_1,d_2,d_3)  = 0 &\\
&  D_{d_4 d_1}^{d_5 d_6} D_{d_2 d_3}^{d_4 d_7} + \mbox{symm}(d_5, d_6,
    d_7) + \mbox{cyclic}(d_1,d_2,d_3) = 0&
\end{eqnarray}
with $ C_{d_1 d_2}^{d_3} $ and $D_{d_1 d_2}^{d_3 d_4}$
antisymmetric in $d_1, d_2$, and $D_{d_1 d_2}^{d_3 d_4}$ symmetric
in $d_3, d_4$. One may also include higher order terms in the bracket
while preserving the existence of a degree decreased by the bracket,
as in the so called spin 4 algebra :
\begin{eqnarray}
[L_{a_1}, L_{a_2}] &=& C_{a_1 a_2}^{a_3} L_{a_3} \\{}
[L_{a_1}, T_{b_1}] &=& C_{a_1 b_1}^{b_2} T_{b_2} \\{}
[L_{a_1}, W_{c_1}] &=& C_{a_1 c_1}^{c_2} W_{c_2} \\{}
[T_{b_1}, T_{b_2}] &=& C_{b_1 b_2}^{a_1} L_{a_1} +  C_{b_1 b_2}^{b_3} T_{b_3} +
                       D_{b_1  b_2}^{a_1 a_2} L_{a_1} L_{a_2} \\{}
[T_{b_1}, W_{c_1}] &=& C_{b_1 c_1}^{b_2} T_{b_2} +
                       D_{b_1 c_1}^{a_1  b_2} L_{a_1} T_{b_2} \\{}
[W_{c_1}, W_{c_2}] &=& C_{c_1 c_2}^{a_1} L_{a_1} +
                       D_{c_1  c_2}^{a_1 a_2} L_{a_1} L_{a_2} +
                       E_{c_1  c_2}^{a_1 a_2 a_3} L_{a_1} L_{a_2} L_{a_3}.
\end{eqnarray}
If one sets $n(L) = 1, n(T) = 3/2, L(W) = 2$, one gets $n([A,B]) \leq
n(A) + n(B)$.

We have checked, using REDUCE, that in both cases the first seven
terms in $\Omega$ are generically non zero.
\section{Conclusion}

We have shown in this paper that polynomial Poisson algebras provide a
rich arena in which the perturbative features of the BRST construction
are perfectly illustrated. We believe this to be of interest because
models of higher rank are rather rare and are usually thought not to
arise in practice. Non polynomial Poisson algebras(e.g. of the type
arising in the study of quantum groups) can also be analyzed along
the same BRST lines and should provide further models of higher rank.

We have not discussed the quantum realization of Poisson algebras, and
whether the nilpotency condition for the BRST generator is maintained
quantum-mechanically. This is a difficult question, which is model-dependent.
Indeed, while the ghost contribution to $\Omega^2$ can be evaluated
independently of the specific form of the $G_a$'s in terms of the
canonical variables $(q^i, p_i)$ (once a representation of the ghost
anticommutation relations is chosen), the ``matter'' contribution to
$\Omega^2$ depends on the ``anomaly'' terms in $[G_a, G_b]$, which, in
turn, depend on the specific form of the $G_a$'s. It would be
interesting to pursue this question further.

\section{Acknowledgments}
We are grateful to Jim Stasheff and Claudio Teitelboim for fruitful
discussions at the early stages of this research.
This work has been supported in part by research funds from FNRS
(Belgium) and by a research contract with the Commission of the
European Communities.


\begin{thebibliography}{10}

\bibitem{Nak:}
M.~Nakamura.
\newblock {\em Prog. Theor. Phys.}, 37:195, 1967.

\bibitem{Pri:}
S.B. Priddy.
\newblock {\em Trans. Amer. Math. Soc.}, 152:39, 1970.

\bibitem{Skl:}
E.K. Sklyanin.
\newblock {\em Funct. Anal. Appl.}, 16:263, 1982.

\bibitem{Zam:}
A.~B. Zamolodchikov.
\newblock {\em Theor. Math. Phys.}, 65:1205, 1985.

\bibitem{FatZam:}
V.A. Fateev and A.~B. Zamolodchikov.
\newblock {\em Nucl. Phys.}, B280:644, 1987.

\bibitem{Oh:}
Y.G. Oh.
\newblock {\em Lett. Math. Phys.}, 12:87, 1986.

\bibitem{TarTakFad:}
V.O. Tarasov, L.A. Takhatadzlupan, and L.D. Faddeev.
\newblock {\em Theor. Math. Phys.}, 57:1059, 1983.

\bibitem{BakMat:}
I~Bakas and P.~Mathieu.
\newblock {\em Phys. Lett.}, 208B:101, 1988.

\bibitem{BhaRam:}
K.~H. Bhaskara and K.~Rama.
\newblock {\em J. Math. Phys.}, 32:2319, 1991.

\bibitem{GraZhe:}
Ya.~I. Granovskii, A.~S. Zhedanov, and I.~M. Lutsenko.
\newblock {\em Theor. Math. Phys.}, 91:474, 1992.

\bibitem{HenTei:QuaGauSys}
M.~Henneaux and C.~Teitelboim.
\newblock {\em Quantization of Gauge Systems}.
\newblock Princeton University Press, 1992.

\bibitem{Sta:}
V.K.A.M. Gugenheim and J.D. Stasheff.
\newblock {\em Bull. Soc. Math. Belgique}, 38:237, 1986;
\newblock J.D. Stasheff.
\newblock {\em Bull. Amer. Soc.},19:287, 1988 and references therein.

\bibitem{Hen:PhyLet}
M.~Henneaux.
\newblock {\em Phys. Lett.}, 120B:75, 1983.

\bibitem{FujKub:}
K.~Fujikawa and J.~Kubo.
\newblock {\em Phys. Lett.}, 199B:75, 1987.

\bibitem{Hue:}
J.~Huebschmann.
\newblock Extensions of Lie algebras.
\newblock Heidelberg preprint, 1989.

\bibitem{Dre:CanExp}
A.~Dresse.
\newblock Canonical form of expressions involving dummy variables.
\newblock Submitted to J. Symb. Comp.

\bibitem{Dre:Imacs}
A.~Dresse.
\newblock Treatment of dummy variables and BRST theory in computer algebra.
\newblock In {\em Proceedings of the IMACS Symposium SC--1993}, 1993.

\bibitem{BurCapDre:}
A.~Burnel, H.~Caprasse, and A.~Dresse.
\newblock in preparation.

\bibitem{ADresse3}
A.~Dresse.
\newblock Ph.D. Thesis, U.L.B. (in preparation).

\bibitem{SchSevNie:QuaBRSChaQua}
K.~Schoutens, A.~Sevrin, and P.~van Nieuwenhuizen.
\newblock {\em Commun. Math. Phys.}, 124:87, 1989.


\end{thebibliography}

\end{document}